\documentclass[twocolumn]{aa}
\usepackage{graphicx}
\begin{document}
   \title{Detection of diffuse TeV gamma-ray emission from the
nearby starburst galaxy NGC 253}

\titlerunning{TeV gamma-ray from NGC 253}

   \author{C.~Itoh\inst{1}\and 
R.~Enomoto\inst{2}\thanks{\email{enomoto@icrr.u-tokyo.ac.jp}}\and
S.~Yanagita \inst{1}\and T.~Yoshida\inst{1}\and 
A.~Asahara \inst{3}\and
G.V.~Bicknell\inst{4}\and
R.W.~Clay\inst{5}\and
P.G.~Edwards\inst{6}\and
S.~Gunji\inst{7}\and
S.~Hara\inst{3,8}\and
T.~Hara\inst{9}\and
T.~Hattori\inst{10}\and
Shin.~Hayashi\inst{11}\and
Sei.~Hayashi\inst{11}\and
S.~Kabuki\inst{2}\and
F.~Kajino\inst{11}\and
H.~Katagiri\inst{2}\and
A.~Kawachi\inst{2}\and
T.~Kifune\inst{12}\and
H.~Kubo \inst{3}\and
J.~Kushida\inst{3,8}\and
Y.~Matsubara\inst{13}\and
Y.~Mizumoto\inst{14}\and
M.~Mori\inst{2}\and
H.~Moro\inst{10}\and
H.~Muraishi\inst{15}\and
Y.~Muraki\inst{13}\and
T.~Naito\inst{9}\and
T.~Nakase\inst{10}\and
D.~Nishida\inst{3}\and
K.~Nishijima\inst{10}\and
K.~Okumura \inst{2}\and
M.~Ohishi\inst{2}\and
J.R.~Patterson\inst{5}\and
R.J.~Protheroe\inst{5}\and
K.~Sakurazawa\inst{8}\and
D.L.~Swaby\inst{5}\and
T.~Tanimori \inst{3}\and
F.~Tokanai\inst{7}\and
K.~Tsuchiya\inst{2}\and
H.~Tsunoo\inst{2}\and
T.~Uchida\inst{2}\and
A.~Watanabe \inst{7}\and
S.~Watanabe\inst{3}\and
T.~Yoshikoshi \inst{16}}
	\authorrunning{Itoh, Enomoto, Yanagita, Yoshida et al.}
   \offprints{R. Enomoto}

   \institute{
Faculty of Science, Ibaraki University,
Mito, Ibaraki 310-8512, Japan 
\and
Institute for Cosmic Ray Research, Univ. of Tokyo, Kashiwa,
Chiba 277-8582, Japan 
\and
Department of Physics, Kyoto University, Sakyo-ku, Kyoto 606-8502, Japan
\and
MSSSO, Australian National University, ACT 2611, Australia
\and
Department of Physics and Math. Physics, University of Adelaide, SA 5005, 
Australia
\and
Institute for Space and Aeronautical Science, Sagamihara, 
Kanagawa 229-8510, Japan
\and
Department of Physics, Yamagata University, Yamagata, Yamagata 990-8560, Japan
\and
Department of Physics, Tokyo Institute of Technology, Meguro-ku, Tokyo 152-8551, Japan
\and
Faculty of Management Information, Yamanashi Gakuin University, 
Kofu,Yamanashi 400-8575, Japan
\and
Department of Physics, Tokai University, Hiratsuka, Kanagawa 259-1292, Japan
\and
Department of Physics, Konan University, 
Hyogo 658-8501, Japan
\and
Faculty of Engineering, Shinshu University, Nagano, Nagano 380-8553, Japan
\and
STE Laboratory, Nagoya University, Nagoya, Aichi 464-8601, Japan
\and
National Astronomical Observatory of Japan, Mitaka, Tokyo 181-8588, Japan
\and
Department of Radiological Sciences, 
Ibaraki Prefectural University of Health Sciences,
Ibaraki 300-0394, Japan
\and
Department of Physics, Osaka City University, Osaka, Osaka 558-8585, Japan}

   \date{Revised 9 Sep. 2002/ Accepted 29 Oct. 2002/ A\&A 396(2002)L1-4}

\abstract{ 
We report the TeV gamma-ray observations of the nearby normal
spiral galaxy NGC~253. 
At a distance of $\sim$2.5~Mpc,
NGC~253 is one of the nearest starburst galaxies.
This relative closeness, coupled with the high star formation rate in
the galaxy, make it a good candidate TeV gamma-ray source.
Observations were carried out in 2000 and 2001 with the
CANGAROO-II 10\,m imaging atmospheric Cerenkov telescope. 
TeV gamma-ray emission is detected at the $\sim 11\sigma$ level with a
flux of 
$(7.8 \pm 2.5)\times 10^{-12} {\rm cm}^{-2} {\rm sec}^{-1}$ 
at energies $>$0.5~TeV. The data indicate that
the emission region is broader than the point spread
function of our telescope. 

}

\maketitle

%
%________________________________________________________________

\section{Introduction}
NGC~253 is a nearby ($\sim$2.5~Mpc) (Vaucouleurs \cite{vau78}) 
starburst galaxy,
in which a high cosmic-ray density and non-thermal emissions are 
expected (Voelk et al. \cite{voe96}).
The large-scale
radio continuum structure of NGC~253 was studied by Hummel et al.\ 
(\cite{hum84}).
Carilli et al.\ (\cite{car92}) found a large, bright
synchrotron-emitting halo extending to more than 10~kpc from the galaxy centre.
Diffuse X-ray emission from the halo has also been detected 
(Cappi et al.\ \cite{cap99};
Strickland et al.\ \cite{str02}).
The OSSE instrument on the {\it Compton Gamma-Ray Observatory} (CGRO)
detected sub-MeV gamma-rays (Bhattacharya et al.\ \cite{bha94}) which 
were attributed to far-infrared photons scattered by the high energy,
synchrotron emitting, electrons  (Goldshmidt \& Raphaeli \cite{gol95}).
The EGRET detector, 
however, gave only upper limits on gamma-ray emission in GeV energies
(Blom et al. \cite{blo99}).

We have observed NGC~253 with the CANGAROO-II telescope between
October 3 and November 18, 2000, and September 20 to November 15, 2001. 
While the detection of diffuse TeV emission from our galaxy is very 
difficult (e.g., Aharonian et al.\ \cite{aha02}), 
observations 
of NGC~253 provide the opportunity to study the distribution
of cosmic rays in galaxies like our own.

\section{Observation and Analysis}

Very high-energy gamma-rays (with energy greater than several hundred GeV)  
interact in the upper atmosphere and produce cascades of secondary photons and 
particles. Secondary electrons and positrons with 
energy greater than the (altitude 
dependent) Cerenkov threshold emit optical photons in 
directions approximately the 
same as that of the incident gamma-ray. The CANGAROO-II 
telescope (Kawachi et al.\ \cite{kaw01}) was constructed near
Woomera, South Australia (136$^\circ$47'E, 31$^\circ$04'S) to 
detect the several nano-second duration
flashes of Cerenkov light. The telescope consists of a 
10\,m reflector and a camera of 552 pixels, 
each a 0.115$^\circ$ square photo-multiplier (PMT).
From the detected Cerenkov 
light images, the directions and energies of the 
incident gamma-rays can be reconstructed.

A total of $\sim$75 hours of observations of 
NGC~253 (``ON'') were made, with a 
similar amount of background (``OFF'') observations. 
Each ON-source run tracked the center of NGC 
253 through its highest elevations, with OFF-source observations 
offset in right
ascension made in order to estimate the background. 
An edge-on galaxy, NGC~253 appears optically 
(based on red Digitized Sky Survey [DSS2] data;
{\tt http://skyview.gsfc.nasa.gov/}),
as a disc $\sim$0.4$^\circ$ by $\sim$0.1$^\circ$ in size.
This is larger than our angular resolution, 
but much smaller than the telescope's field 
of view (FOV) of 1.84$^\circ$ $\times$ 1.84$^\circ$. 
After selecting data taken at high elevation angles 
($>70^\circ$) in good weather conditions, a total of 2959 min.\
ON-source and 2417 min.\ 
OFF-source data remained for further analysis.

\begin{figure}[htbp]
  \centering 
    \includegraphics[width=8cm]{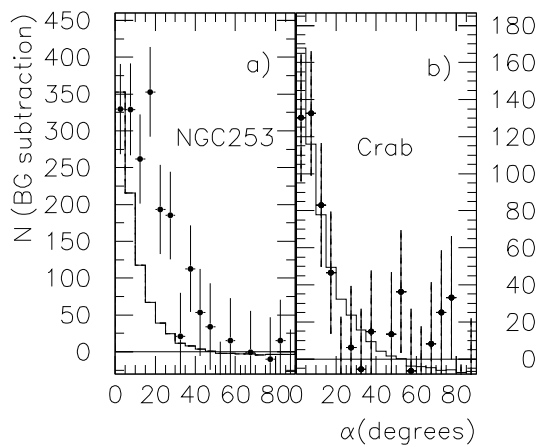} 
\caption{
Distributions of the image orientation angle ($\alpha$): a) for NGC~253
and b) for the Crab. The points with error bars were obtained by
subtracting the normalized off-source data from the on-source data.
The ratio of
events in the higher $\alpha$ ($>$ 30$^\circ$)
regions for on- and off-source data was used as
the normalization factor, which agreed with the ratio of observation times to
within 2\%.
The histograms were obtained
from Monte Carlo simulations of gamma-rays from a point-source.  }
\label{fig1}
\end{figure}

First, ``cleaning'' cuts  
were applied to the camera images, requiring pixel pulse-heights 
of greater than $\sim$3.3~photoelectrons, 
Cerenkov photon arrival times within 40~nsec, and clusters of at
least four adjacent triggered pixels in each event.
After these pre-selections, we carried out a shower image analysis using the
standard set of image parameters, distance, length, width, 
and $\alpha$ (Hillas \cite{hil85}), combining length and 
width (after an initial distance cut) to assign likelihoods 
to each event (Enomoto et al.\ \cite{eno02_1,eno02_2}). 
The  likelihood for both a gamma-ray origin and cosmic-ray 
proton origin were calculated.
The cut 
to reject background events was based on the ratio of 
these two likelihoods. After these 
cuts, the image orientation angle ($\alpha$) was plotted. 
A gamma-ray signal appears as an excess at 
low $\alpha$ after the normalized OFF-source $\alpha$ distribution 
is subtracted from the ON-source 
distribution. As shown in Fig.~\ref{fig1}a, an excess of events with 
$\alpha < 30^\circ$ is clearly observed 
for NGC~253. 
The number of gamma-ray-like events is 
1652$\pm$149 (11.1$\sigma$), corresponding to 
0.56~gamma-rays per min. 

The energy scale (the conversion factor from ADC value to
number of photoelectrons)
was determined by adjusting the cosmic ray
event rate and the relation between the total ADC counts
and the total number of hit pixels.
This estimation also agreed with the calculated 
single trigger rate for pixel due to
the Night Sky Background (NSB).
We obtained the threshold energy for the gamma-ray
detection to be $\sim$ 500~GeV for a $E^{-2.5}$ spectrum
and $\sim$ 400~GeV for a $E^{-3.0}$ spectrum,
after pre-selection cuts and also the same values after image parameter cut.
These values were obtained from the peak positions of the effective area
multiplied by power law energy dependences.

The same analysis procedure was applied to Crab nebula data from 
observations in November and December 2000 as a check. 
Approximately 10~hours of
good data were obtained. The resulting $\alpha$ plot, 
shown in Fig.~\ref{fig1}b, has an excess of 393$\pm$59 
events (6.7$\sigma$). The solid histograms are the 
Monte Carlo predictions for the $\alpha$ 
distribution in the case of a point source. 
Note that the Crab 
observation was carried out at large zenith angles, 
around 56$^\circ$. At such large 
zenith angles, the $\alpha$ distribution deteriorates 
due to the shrinkage of Cerenkov 
images. In order to compensate for this, we applied 
tighter cuts in the Crab 
analysis. As a result, the $\alpha$ distribution for 
Crab is slightly wider than that for 
NGC~253 (which is observed near the zenith). The experimental result, 
however, agrees with the Monte Carlo prediction in case of the Crab 
observation.
For NGC 253, the $\alpha$ distribution is broader than the 
point spread function (PSF) by a factor 
of two. Our Monte Carlo simulations predicted that 73\% of 
events from a point source 
would have $\alpha < 15^\circ$, whereas in fact 56\% of events 
were contained within this range.

\section{Results}

\begin{figure}[htbp]
  \centering 
\includegraphics[width=6cm]{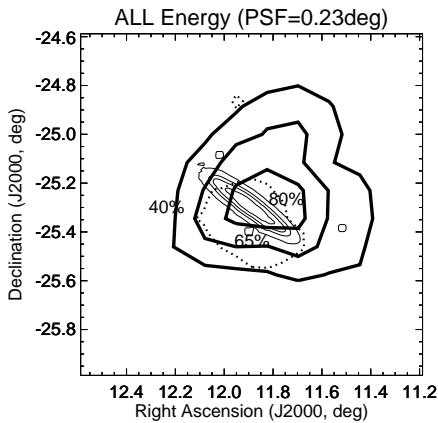}
\caption{
Profile of the emission around NGC~253. 
The solid thick contours were obtained from our observations. 
This was made from the distribution of the detection significance determined at 
each location from the differences in the $\alpha$ plots (ON- minus OFF-source 
histogram) divided by the statistical errors. 
The solid thin 
contours are the DSS2 (red) data. 
The dotted 
contour was obtained with the CANGAROO-II telescope for the Crab. 
}  
    \label{fig2}
\end{figure}

The spatial distribution of the gamma-rays from NGC~253 was 
studied using the 
so-called significance map. The thick solid contours in 
Fig.~\ref{fig2} are our results. 
The 65\% 
confidence level contour, which is roughly one standard 
deviation of our angular
resolution, is shown and compared with an optical image 
from DSS2, which is shown by the solid 
thin contours. The size of the 
TeV emission region is of the same order as the optical image, 
or larger. The dotted 
contour is the 65\%-contour obtained by the Crab observation. 
It is consistent with the 
Monte Carlo prediction of $0.25^\circ$.
That of the near-zenith observations was estimated to be $0.23^\circ$. 
In order to estimate the 
telescope pointing accuracy, we analyzed bright stars 
(optical magnitude from 5 to 4) in 
other observations at various elevation angles, 
as the stars in the NGC~253 field of view 
are fainter than our detection limit. We conclude that 
the pointing uncertainty was 
within $0.1^\circ$, which is less than the pixel size. 
The center of gamma-ray emission from the Crab observation 
agreed within this accuracy. 
The gamma-ray acceptance of the telescope gradually decreases beyond 
$0.3^\circ$ from the center of the field 
of view (dropping to half at $0.8^\circ$), and so we are unable to 
make a more detailed morphological 
study. It is clear, however, that the observed 
gamma-rays distribution from NGC 253 is inconsistent with 
emission from a point source.

In order to derive the flux level, we estimated 
the acceptances (effective areas) using Monte 
Carlo simulations (Enomoto et al.\ \cite{eno02_1}). 
We compared
our effective area with that of Whipple
(Fig. 6 in Mohanty et al.\ \cite{moh98}),
which are these after clustering and distance cut,
i.e., before image parameter cut.
Our effective area agreed with Whipple including the energy dependences.
According to Fegan (\cite{feg96}), the threshold
of Whipple telescope was the same as ours.
The sensitivity for the CANGAROO telescope was represented
by that for Crab-type source, i.e., a point source with Crab 
intensity at zenith.
We calculated its acceptance and assumed the same background level,
with the cut of $\alpha < 15$ degrees.
It corresponds to $\sim$8$\sigma$ detection per hour, slightly
higher than that of Whipple (Reynolds et al. \cite{rey93}). 
A difference appeared after the shape parameter cut, 
i.e., likelihood analysis.
Our acceptance for the likelihood cut was as high as 86\%. On the other hands,
the ``supercut" reduces it to 50\% level (Fegan \cite{feg96}: Table II).
The both background levels are the same.
Note that our background level depends on
NSB, artificial light background and etc., 
because we used the experimental OFF sample as the background
sample for Probability Density Functions.

\begin{figure}[htbp]
  \centering 
 \includegraphics[width=9cm]{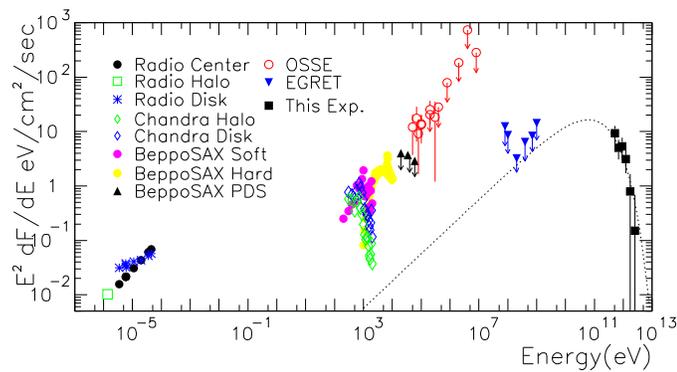} 
\caption{
Multi-band spectrum of NGC 253. The black squares were 
obtained by this experiment. 
When we constrain the flux to 3/4 of the EGRET upper limit
(Blom et al.\ \cite{blo99}) at 0.2~GeV using the 
function in the text, a flux proportional to 
$E^{-1.5}e^{-\sqrt{E} /(0.25\pm 0.01)}$ was obtained with 
$\chi^2$/DOF=1.8/5 (the dotted line). 
The X-ray data for CHANDRA and 
BeppoSAX soft component were corrected for photo-absorption in the Galaxy,
but no correction was applied to the 
BeppoSAX hard X-ray data.
No correction was made for photo-absorption inside NGC~253 itself. 
We note that the 
BeppoSAX, OSSE, and EGRET data were not able to spatially resolve
NGC~253, in contrast to the other data.
}
\label{fig3}
\end{figure}

The systematic uncertainty was calculated on a bin-by-bin
basis by changing the cut values on the likelihood from 0.1 to 0.6 in 
steps of 0.1. The uncertainties are typically $\sim$30\%.
The energy scale uncertainties
due to the mirror reflectivity, mirror segment distortions,
and Mie scatterings
were estimated to be 15\% (point to point) and 20\% (overall). The energy
resolution was estimated to be 35\% on an event by event basis. 
The energy resolution is bad because we could not assume source
point event by event basis, i.e., we could not correct it using
core distance. The spill over effect, therefore, is considered to
be large for such soft energy spectrum.
In order to derive the spectrum of NGC~253 self-consistently,
we adopted the iteration method.
At first we used an acceptance calculated by
the simulation with a $E^{-2.5}$ spectrum and
we obtained the index of $-$3.7$\pm$0.3.  We then iteratively
used this fitted value in the simulations to re-derive the spectrum.
This process rapidly converged at an index of $-$3.75$\pm$0.27.
The same iteration procedure was carried out using a function
with cutoff. It also showed a good convergence.
The differential
fluxes obtained in 2000 and 2001 agreed within these errors. 
A check was made using Crab nebula data 
(with an energy threshold 
of $\sim$2~TeV) and the derived flux agreed with 
previous measurements within the 
statistical errors ($\sim 20\%$)
(Tanimori et al.\ \cite{tan98}; Aharonian et al.\ \cite{aha00}). 
Also, the cosmic ray 
flux calculated from the background event 
rate was checked and found to match the expected rate within 10\%. 
A more detailed description of 
the analysis will be presented elsewhere (Itoh et al.\ \cite{ito02_1}). 
The differential fluxes are plotted in the 
spectral energy distribution (SED) shown in Fig.~\ref{fig3}, 
together with measurements from 
other energy bands (Carilli et al.\ \cite{car92}; Hummel et al.\ \cite{hum84};
Cappi et al.\ \cite{cap99};
Strickland et al.\ \cite{str02}; Bhattacharya et al. \cite{bha94};
Blom et al.\ \cite{blo99}). 

A simple power-law fit to the TeV spectrum of NGC~253 deviates greatly 
from the EGRET measurements at GeV energies (Blom et al.\ \cite{blo99}): 
a turn-over below the TeV region 
clearly exists. We fitted a power law with an exponential cut off, 
fixing the power-law 
index to $-$1.5, i.e., an inverse Compton--like spectrum 
(where the exponential term 
is exp$(-\sqrt{E}/a)$, with $a$ a free parameter). 
The resulting improved fits are shown by 
the dotted curve in Fig.~\ref{fig3}. 
The minimum power-law index providing an acceptable fit
(based on the $\chi^2$ value) was $-$1.8 (2$\sigma$ limit). 
If the observed gamma-rays 
are produced by inverse Compton 
scattering of cosmic microwave background (CMB) photons, 
then the fitted spectrum 
suggests that the maximum energy of the parent 
electrons is around several~TeV. 
The integral flux is estimated to be 
$(7.8\pm 2.5) \times 10^{-12} {\rm cm}^{-2} {\rm s}^{-1}$
at energies greater than 0.52~TeV.

\section{Discussions}

The observations reported here are the first detection 
of TeV gamma-rays from a normal 
spiral galaxy (other than our own), and reveal the 
emission to be spatially extended and 
temporally steady. 
NGC~253 has been observed over 
a range of photon energies, as depicted in 
Fig.~\ref{fig3}, however the SED is more difficult to interpret. 
The emission of 50--200~keV 
photons observed by OSSE (Bhattacharya et al.\ \cite{bha94}) was interpreted 
as being inverse Compton scattering of
 far-infrared photons from dust in the central region 
(Goldshmidt \& Raphaeli \cite{gol95}). 
A simple extrapolation to higher 
energies exceeds the EGRET 2$\sigma$ upper limit (Blom et al.\ \cite{blo99}). 
Clearly, it is not 
possible to explain the OSSE 
and 
CANGAROO-II observations and the EGRET upper limits by invoking inverse 
Compton emission from a single population of electrons. 
The very large radio halo, 
extending over $\sim$10~kpc (Hummel et al.\ \cite{hum84};
Carilli et al.\ \cite{car92})
and up to X-ray 
energies (Cappi et al.\ \cite{cap99}), suggests the existence of a 
population of very high-energy cosmic rays which 
may be responsible for the inverse 
Compton production of TeV gamma-rays (Itoh et al.\ \cite{ito02_2}), 
quite separate from sources concentrated near the 
centre of the galaxy which emit the majority of the photons observed by OSSE. 

NGC~253 is the first of a new class of object to be detected at TeV energies.  
The extended and steady nature of the emission makes it clear 
that the TeV gamma-rays are 
produced in a different environment than that of 
previously reported extragalactic 
sources of the active galactic nucleus class. Studies of NGC~253 will, 
like those of the LMC at GeV 
energies (Sreekumar et al.\ \cite{sre92}), provide the opportunity to 
learn more about the cosmic ray environment in 
galaxies like our own.

\begin{acknowledgements}

%This work was supported by a Grant-in-Aid for Scientific Research by
%the Japan Ministry of Education, Science, Sports and Culture,
%Australian Research Council, and Sasagawa Scientific Research Grant
%from the Japan Science Society. 
We thank Prof. T. G. Tsuru for
suggestions.
%The Digitized Sky Survey was produced at the Space Telescope Science
%Institute under U.S. Government grant NAG W-2166. The images of these
%surveys are based on photographic data obtained using the Oschin
%Schmidt Telescope on Palomar Mountain and the UK Schmidt
%Telescope. The plates were processed into the present compressed
%digital form with the permission of these institutions.

\end{acknowledgements}

\end{document}